\documentstyle[epsf,floats,prb,aps]{revtex}
\begin{document}
\draft
\title{Onset of magnetism in B2 transition metals aluminides }
\author{N.~I.~Kulikov,}
\address{Institute of High Pressure Physics,
Troitsk, Moscow region, 142092, Russia}
\author{A.~V.~Postnikov, G.~Borstel, and J.~Braun}
\address{Fachbereich Physik, Universit\"at Osnabr\"uck,
D-49069 Osnabr\"uck, Germany}
\date{Received December 23, 1997; in final form October 14, 1998}
\twocolumn[\hsize\textwidth\columnwidth\hsize\csname@twocolumnfalse\endcsname
\maketitle
\begin{abstract}
{\em Ab initio} calculation results for the electronic structure
of disordered bcc Fe$_x$Al$_{1-x}$ (0.4 $<x<$ 0.75),
Co$_x$Al$_{1-x}$ and Ni$_x$Al$_{1-x}$ ($x$=0.4; 0.5; 0.6)
alloys near the 1:1 stoichiometry, as well as of the ordered B2
(FeAl, CoAl, NiAl) phases with point defects are presented.
The calculations were performed using the coherent potential
approximation within the Korringa-Kohn-Rostoker method (KKR-CPA)
for the disordered case and the tight-binding linear muffin-tin orbital
(TB-LMTO) method for the intermetallic compounds.
We studied in particular the onset of magnetism in Fe--Al and Co--Al
systems as a function of the defect structure. We found the appearance
of large local magnetic moments associated with the transition metal (TM)
antisite defect in FeAl and CoAl compounds, in agreement
with the experimental findings. Moreover, we found that any vacancies
on both sublattices enhance the magnetic moments via reducing
the charge transfer to a TM atom. Disordered Fe--Al alloys are
ferromagnetically ordered for the whole range of composition studied,
whereas Co--Al becomes magnetic only for Co concentration $\geq$ 0.5.
\end{abstract}
\pacs{
75.20.En 
75.50.Bb 
71.23.-k 
71.15.Mb 
}
]

\section{Introduction}
The intermetallic compounds FeAl, CoAl and NiAl attract considerable
attention due to their uncommon properties. For example,
a high melting point makes them attractive as promising
high-temperature aerospace materials. Furthermore, an unusual magnetic
behavior, depending on temperature and concentration, has been observed.
The $\beta$-phase of these aluminides crystallizes in the B2 (CsCl) structure
and persists over 45--58 at.\% in the case of Co;
$\beta$-phases of Ni and Fe aluminides are stable over
a broader range of composition. It is known that these compounds have
high concentrations of point defects, and moreover exhibit specific
defect structures known as triple defects.\cite{W}
Predominant are apparently combinations of so-called
antistructure (AS) atoms -- usually transition metal (TM) atoms
substituting Al (in the Al-poor region of the $\beta$ phase),
and vacancies on transition-metal sites (in the Al-rich region).

In spite of their structural similarity, these materials exhibit different
mechanical and magnetic properties. At low temperatures (below 10 K),
the nearly equiatomic alloy Co$_{0.506}$Al$_{0.494}$
shows a temperature dependence of the magnetic susceptibility
which coincides neither with the Curie nor with the Curie-Weiss law.\cite{LB}
There is a common agreement in the literature that the perfectly
ordered compound CoAl is nonmagnetic,\cite{SCF,AG} but already
a small disorder leads to the onset of local moments
(see, e.g., Ref.\onlinecite{SZD} and references therein). Also it is widely
believed that the Co-AS atoms are responsible for the magnetic properties of
this compound: several authors tried to deduce from magnetic
measurements the moment per Co-AS atom which is associated with an
effective moment of ~5$\mu_{\mbox{\tiny B}}$.\cite{LB,SCF} It is necessary
to note that above 60 at.\% Co, a permanent magnetization
appears at low temperatures.\cite{SCF}
For the NiAl case, no magnetic moment has been observed for any concentrations.
Earlier {\it ab initio} calculations for ordered FeAl\cite{SSKPD,ZF} reveal
a moderate magnetic moment of 0.6--0.7 $\mu_{\mbox{\tiny B}}$ at the Fe site
and a ferromagnetic ordering. Ordered CoAl and NiAl were
found in earlier {\it ab initio} calculations
to be nonmagnetic, which is in agreement with the measurements.

The experimental magnetic phase diagram shows that FeAl near stoichiometry
is in a spin-glass phase,\cite{SW} but in cold-worked disordered
Fe-Al alloys the saturation magnetization persists in
both stoichiometric FeAl and Al-rich alloys.\cite{BHM}

The concentration dependence of the lattice parameter for the
$\beta$-phase of aluminides is also not trivial. Typically,
one should expect some slight deviations from the Vegard's rule,
i.e. nearly linear variation with composition.
In CoAl and NiAl however, the lattice spacing has
the maximum at approximately equiatomic concentration
of the alloy.\cite{BZSG,Co}
In FeAl, the lattice constant is nearly constant over a wide range of
composition from 30 to 51 at.\% Al.\cite{BHM}
This fact has so far no theoretical explanation. Nevertheless one may expect
such behavior to be due to the lattice dilatation,
depending on the concentration of vacancies and antisite (AS) atoms,
that in its turn is controlled by composition and heat treatment.

The microscopic understanding of these phenomena can be obtained from
electronic structure calculations. Actually the electronic structure of
magnetic TM aluminides with stoichiometric composition has been already
studied for many times by various methods in different approximations.
In the earliest calculations within the Korringa-Kohn-Rostoker (KKR)\cite{MWJ}
and modified KKR methods\cite{MBZ} the problem of filling up
of the TM $d$-bands by Al $p$-electrons was discussed
in detail and the charge transfer was shown from an Al to a TM site.
The possibility of magnetic ordering was studied for the Ni--Al system
by the augmented spherical wave (ASW) method\cite{HK}. Hereby only
for Ni$_3$Al a ferromagnetic ordering was found, with a value of
the Ni magnetic moment by an order of magnitude less than
in pure Ni. Optical properties of TM aluminides were studied
by the linear muffin-tin orbital (LMTO) method\cite{KK}
and by the linearized-augmented-plane-wave (LAPW) method\cite{KHL}
for NiAl and CoAl. The experimental absorption maxima
were found to be correlated to the band structure of these alloys.\cite{ST}
Further details of the electronic structure of these
compounds were studied by comparing the data from electron energy-loss
spectroscopy with theoretical investigations performed
by the LMTO method.\cite{BGTH} The trends in the chemical bonding and
the phase stability of transition metal aluminides with equiatomic
composition have been studied by the full-potential linearized augmented
plane-wave (FLAPW) method.\cite{ZF} 
A review of electronic structure calculation results, along with
band structures, densities of states and Fermi surfaces of many
TM aluminides can be found in Ref.~\cite{Si94}.
A recent study using the full-potential linearized augmented
Slater-type orbital method\cite{Wa98} reports the formation
energies and equilibrium volume of many $3d$ aluminides.

The equation of state and all
zero-pressure elastic moduli for CoAl have been calculated\cite{MOPK} also
using the FLAPW method. In addition, cohesive, electronic and
magnetic properties of the transition metal aluminides have been calculated
using the tight-binding linear muffin-tin orbital (TB-LMTO) method.\cite{SSKPD}
Here it was found that only FeAl retains a magnetic moment,
which has been mentioned above.
These findings coincide with the results of earlier LMTO calculations
for NiAl and FeAl intermetallic compounds\cite{MOJF} and with
{\em ab initio} pseudopotential calculations\cite{OR} for CoAl.

There are relatively few calculations aimed at the study of defects'
influence on the electronic and magnetic structure of TM
aluminides. {\em Ab initio } electronic structure
calculations for point defects in CoAl have been performed
by Stefanou {\it et al.}\cite{SZD} within the KKR-Green's function method.
Herein the perturbation of the potential was included at the sites
which are nearest to the substitutional impurity.
Earlier, Koenig {\it et al.}\cite{KSK} performed Green's function
calculations for vacancies in B2 aluminides by the
LMTO-Green's function method, but allowing only the potential
at the impurity site to be perturbed.
The LMTO method has been applied to study the electronic
structure of AS defects in FeAl where the point defect was
modelled by suitably chosen supercells.\cite{GF}
Finally, the LMTO-CPA technique has been used for the calculation
of electronic and thermodynamic properties of disordered
NiAl alloys.\cite{AVRK}
In Ref.\onlinecite{BDKJA}, the LMTO-CPA calculation results
are presented for discussing the order-disorder transition in FeAl alloys.

The supercell approach has been used in order to study the antiphase
boundary in NiAl and FeAl\cite{LXF,Fu} as well as point defects
in these aluminides.\cite{Fu}
Very interesting results for the magnetism of Co-doped NiAl by FLAPW
supercell calculation have been obtained by Singh,\cite{Si92} and we
compare our findings with these results.

Although many studies have been performed up to now, the aspects of
magnetism seen as a function of concentration remains
not completely clear in these compounds. The purpose of this paper
is therefore to investigate the onset of magnetism in
the $\beta$-phase of ferromagnetic TM (Fe, Co, Ni)
aluminides. First-principles total energy calculations have been performed
in order to determine the magnetic phase stability of these alloys.
The calculations are based on the density-functional (DF) theory.
We have applied the TB-LMTO method in order to solve the DF equations
in the case of near-stoichiometric ordered compounds with points defects.
The KKR-CPA has been used for disordered substitutional alloys.
We also present here the results of our
supercell simulations for AS atoms and vacancies for both sublattices
of the ordered compounds and compare them with the results for completely
disordered alloys in order to clarify the physical reasons for the
magnetic instability in these systems.

This paper is organized as follows: Sec. II is devoted to the description
of our calculational methods, in Sec. III we discuss the results of our
TB-LMTO calculations for the point defects. In Sec. IV we present
the results for disordered alloys, concentrating on Co-Al as the most
interesting one. A summary is given in Sec. V.

\section{Methods of calculation}
\label{methods}
The conventional {\em linear muffin-tin orbital} method in the
{\em atomic spheres approximation} (LMTO-ASA) as well as its
tight-binding formulation (TB-LMTO) have been well
described in the literature.\cite{OKA,AJ}
Here we present only the relevant computational details.

We have used equal and space-filling atomic sphere radii on Al
and TM sites in all calculations, scaling them with the lattice
constant when studying the volume trends.
According to the experience of previous studies
(see, e.g. Ref.~\onlinecite{MM}),
this choice gives the best description of electronic
densities for these systems in the ASA.
Basis functions up to $l=2$ were included for both constituent atoms
explicitly, and the $l=3$ functions via the downfolding
procedure.\cite{DwnF} Combined corrections have been included
throughout, and the effect of non-local corrections
(after Langreth and Mehl, Ref.~\onlinecite{LM}) to the
local density approximation (LDA), i.e. compared with the
exchange-correlation potential by von Barth and Hedin,\cite{vBH}
has been specially addressed and discussed. 
Since the advent of schemes incorporating the non-local functional
forms of the exchange-correlation energy in terms of electron density,
they have been applied to many solid-state problems, including
magnetic ones. Whereas the non-local corrections were shown
to improve the LDA-based description in many cases (providing,
e.g., the correct total energy hierarchies of magnetic phases),
one should keep in mind that they tend to underestimate the bonding
and hence to lead to an overestimated equilibrium volume
(otherwise typically underestimated within the LDA). Moreover,
the non-local corrections favor larger magnetic polarization.
In some systems, where the interplay of magnetic properties and volume
is crucial, one must be careful in applying the non-local-corrected
schemes. In the present study, we in many cases refer to the results
obtained both within the LDA and incorporating the non-local corrections.
Taken together, they provide reasonable error bars for the properties
obtainable from a first-principle calculation scheme.

The tetrahedron method integration\cite{Blo-k} have been performed on the
$16\!\times\!16\!\times\!16$ mesh over the Brillouin zone (BZ),
so that the total energy values cited below were converged
against the $k$-mesh enhancement within relevant digits,
important for the study of the volume trends.

The electronic structure of disordered Co--Al alloys have been calculated
using the fast KKR-CPA technique.\cite{TK}
The KKR-CPA method is based on the multiple scattering theory.
In this case the electronic Green's function has to be calculated
self-consistently with the scattering path operator
for an atom embedded into the ordered lattice
of effective scatterers. In addition, the potential for each
constituent has to be brought to self-consistency in the CPA sense.
Therefore in KKR-CPA we have two self-consistent procedures,
that leads to a number of numerical difficulties.
An effective solution of this problem has been proposed
in Ref.~\onlinecite{TK}. Specifically,
for the BZ integration the tetrahedron scheme with complex energies
is applied, and in the CPA loop the extrapolation approach with only
few iterations is introduced in order to solve the CPA equation
on the rectangular contour in the complex energy plane.
As a rule, we used in the present work the same parameters for the energy
integration contour as described in Ref.~\onlinecite{TK}.
The necessarily minimal basis set included $s$, $p$, $d$ and $f$
orbitals for both TM and Al atoms, at which all space-filling atomic
spheres had the same size. In this sense, the calculation setup
was close to that used in TB-LMTO calculations for the ordered structures.
Another similarity is that no frozen core approximation has been
used, i.e. both schemes are all-electron ones. The difference is,
TB-LMTO calculations were scalar-relativistic ones whereas KKR-CPA
calculations were nonrelativistic, but this difference seems to have
only negligible effect on structural and magnetic properties
of TM aluminides.
Also, the effects of non-local density functionals have not been
studied in our KKR-CPA calculations. The exchange-correlation
has been treated there in the LDA
according to the Hedin--Lundqvist prescription\cite{HL}
for non-magnetic systems and with spin scaling
after von Barth and Hedin\cite{vBH} for magnetic alloys.

\begin{figure}[th]
\epsfxsize=8.6cm
\centerline{\epsfbox{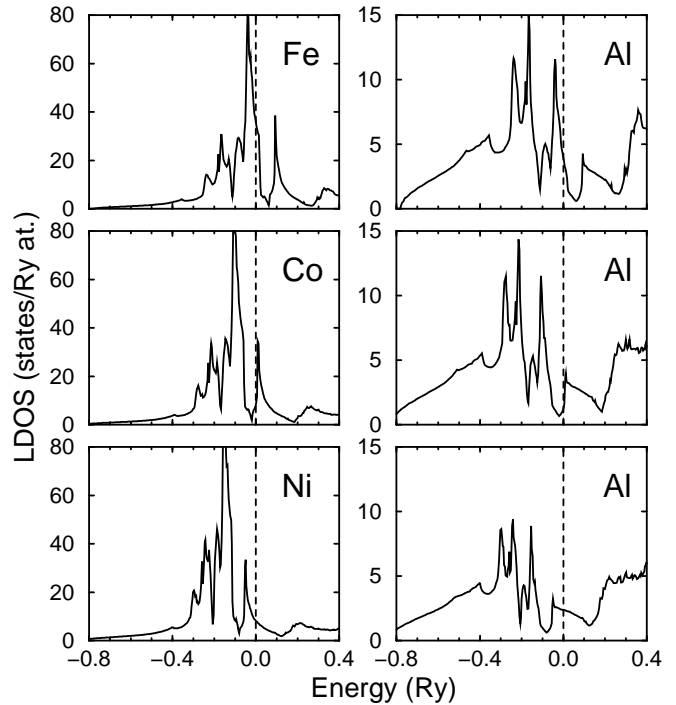}}
\vspace*{0.5cm}
\caption{Local DOS of components in the non-magnetic case for ordered
FeAl, CoAl and NiAl as calculated by TB-LMTO. The dashed line indicates
the Fermi level.}
\label{Single_DOS}
\end{figure}

In contrast to the LMTO method where all energy eigenvalues
are explicitly provided by matrix diagonalization
and the application of the tetrahedron method is straightforward,
the BZ integration is more involved in the KKR method.
In order to trace the {\it a priori} unknown number of poles
in the energy-dependent determinant, we have divided
the irreducible part of the BZ into 1024 equivalent
tetrahedra by an uniformly distributed mesh of $k$-points.
While examining the tetrahedra one by one, we have divided
those where the phase of the integrand changes by more than $\pi/3$
over the volume into eight smaller tetrahedra, etc.
This subdivision allows us to reduce the number of $k$-points
necessary to obtain the relevant precision by a factor of 2--2.5.

In consequence, 10--20 iterations using the modified Broyden scheme\cite{VL}
for accelerating the convergence have been sufficient in the potential loop,
at which the CPA self-consistency being achieved inside each iteration
for the potential. The charge self-consistency has been
used as the convergency criterion, and the iterations have been performed
until the electronic density differences were stable to within 10$^{-7}$.
Apart from conventional total energy and electronic structure calculations
within the CPA, we have used the fixed-spin-moment (FSM) procedure\cite{WMKS}
in order to analyze the volume/magnetic moment total energy dependence
in Co--Al alloys.

\section{Electronic structure of point defects in
C\lowercase{o}A\lowercase{l} and F\lowercase{e}A\lowercase{l} compounds}
The above cited previous calculations of ordered FeAl,
CoAl and NiAl compounds have shown
that their electronic structures are rather similar and
related roughly by a rigid-band shift, which is depending on the number
of conduction electrons and hence on the placement of the Fermi energy.
This is illustrated by the electronic
densities of states (DOS) for these materials, which are shown in
Fig.~\ref{Single_DOS}. The ferromagnetism of FeAl is perfectly
consistent with the Stoner model, as the Fermi level
crosses the highest peak of a pronouncedly two-peak DOS distribution.
For CoAl and NiAl, the Fermi level is outside of the region which
is favorable for the onset of ferromagnetism.

Singh\cite{Si92} emphasized the existence of a peak in the ``rigid'' DOS
between the positions of the Fermi level in CoAl and NiAl.
One can expect that tuning the number of conduction electrons,
e.g. doping NiAl with Co, would favor the appearance of
the magnetic order. In Ref.~\onlinecite{Si92}, for example,
the effect of substitutional Co in NiAl has been investigated
in a sequence of supercell calculations
with the FLAPW method. An instability against ferromagnetism was found
for Co$_{0.25}$Ni$_{0.75}$Al and Co$_{0.125}$Ni$_{0.875}$Al compositions.
Apart from this, strong deviations from the rigid-band behavior
were found for doped systems, implicating that the onset of
magnetism in these systems may also be strongly dependent
on the short-range order.

In the present study, we are primarily interested in the influence
of local defects (vacancies and antisites), 

\begin{figure}[h]
\epsfxsize=8.6cm
\centerline{\epsfbox{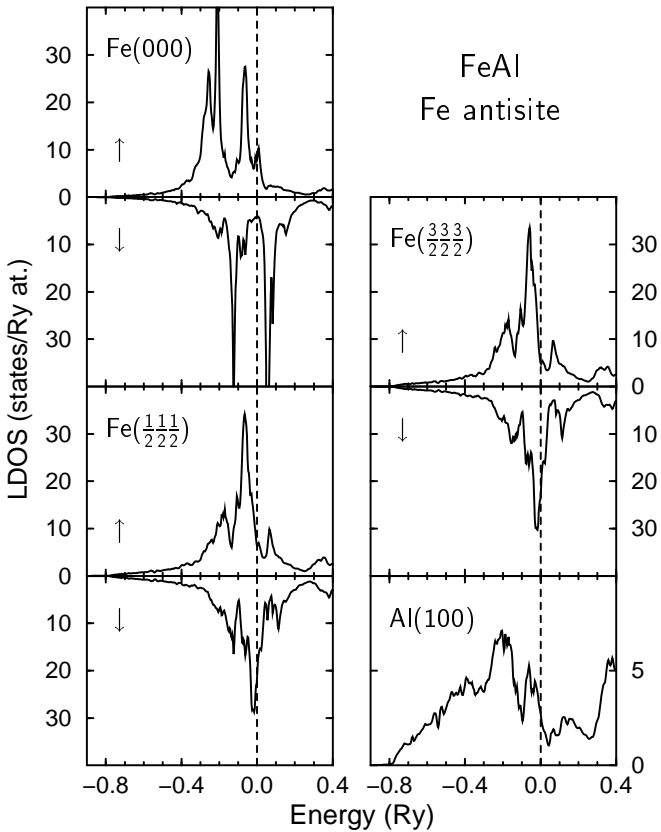}}
\vspace*{0.5cm}
\caption{Local DOS at the AS Fe defect and its several neighbors
in FeAl as calculated by TB-LMTO. The dashed line indicates the Fermi level.}
\label{Fe-AS_DOS}
\end{figure}

\begin{figure}[th]
\epsfxsize=8.6cm
\centerline{\epsfbox{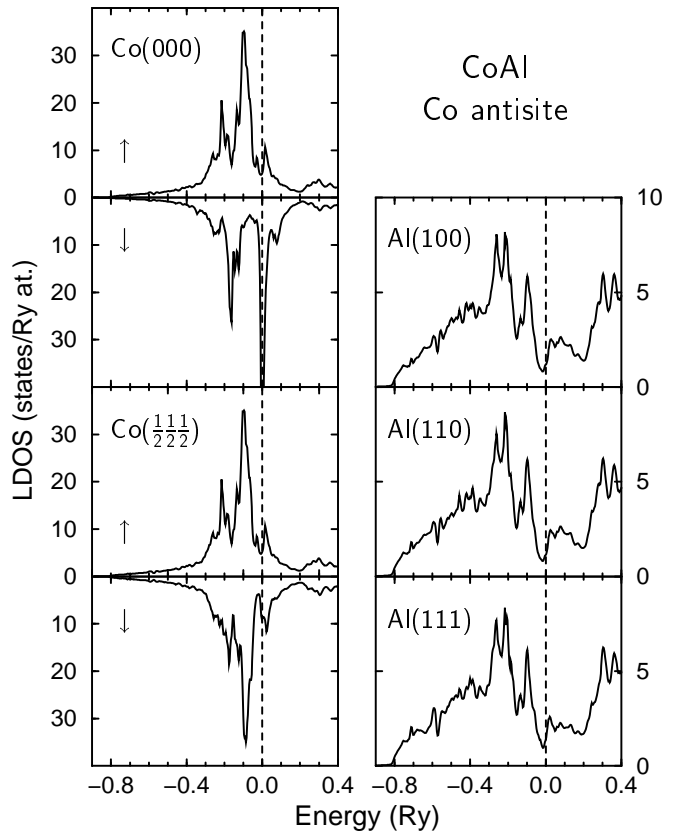}}
\vspace*{0.5cm}
\caption{Local DOS at the AS Co defect and its several neighbors
in CoAl as calculated by TB-LMTO. The dashed line indicates the Fermi level.}
\label{Co-AS_DOS}
\end{figure}

\noindent
which we have introduced
on both sublattices in FeAl as well as in CoAl. In order to suppress the
defect-defect interactions, quite large supercells have been used.
For examining the energetics (i.e., optimizing the lattice spacing
for different defect configurations), we have used supercells
with translation vectors $(0,2a,2a)$, $(2a,0,2a)$, $(2a,2a,0)$
of the underlying cubic lattice, i.e. including 32 atoms.
For reasons of controlling and for the thorough study of the spatial charge
and magnetic moment distribution, we have performed several calculations
with twice larger supercells, spanned up by the
translation vectors $(-2a,2a,2a)$, $(2a,-2a,2a)$, $(2a,2a,-2a)$.
In the last case, the next translated defect appears only as
the 9th nearest neighbor within the corresponding sublattice.
Since the calculations are fully self-consistent, they do probably
provide a more accurate description of the screening of defects
as was previously obtained with the use of KKR--Green's function
method,\cite{SZD} where the potentials only at the defect
site and its nearest neighbors were allowed to deviate from
the perfect-crystal values.

\begin{table*}[ht]
\caption{
Uncompensated number of electrons in the atomic sphere $Q$
and local magnetic moments $m$
in perfect ordered FeAl, CoAl and over shells of atoms
around point defects in these alloys. The results are
from 32-atom supercell calculations
with nonlocal exchange-correlation potential and $a$=5.40 a.u.
}
\begin{tabular}{ccddddddd}
Distance,    & Atom & \multicolumn{4}{c}{FeAl} & \multicolumn{3}{c}{CoAl} \\
\cline{3-6} \cline{7-9}
units of $a$ &      & $Q$ & $m, \mu_{\mbox{B}}$ &
                      $Q$ & $m, \mu_{\mbox{B}}$ &
                      $Q$ & $m, \mu_{\mbox{B}}$ & $Q$ \\
\hline
\multicolumn{2}{c}{Sites in perfect bulk:}&
\multicolumn{2}{c}{Fe}&\multicolumn{2}{c}{Al}&
\multicolumn{2}{c}{Co}&Al\\
       &                   & $-$0.240 &    0.761 &    0.240 & $-$0.036 &
                             $-$0.330 &    0     &    0.330 \\
\hline
 & &\multicolumn{2}{c}{Fe-antisite}&\multicolumn{2}{c}{Al-vacancy}&
    \multicolumn{2}{c}{Co-antisite}&Al-vacancy\\
 0.000 & \mbox{TM or Vac.} & $-$0.021 &    2.239 & $-$1.089 &    0.029 &
                             $-$0.061 &    1.526 & $-$1.047 \\
 0.866 & \mbox{TM}         & $-$0.204 &    0.759 & $-$0.019 &    1.406 &
                             $-$0.280 &    0.143 & $-$0.131 \\
 1.658 & \mbox{  }         & $-$0.247 &    0.648 & $-$0.284 &    0.534 &
                             $-$0.337 &    0.027 & $-$0.358 \\
 1.000 & \mbox{Al}         &    0.254 & $-$0.039 &    0.288 & $-$0.052 &
                                0.350 & $-$0.011 &    0.387 \\
 1.414 & \mbox{  }         &    0.235 & $-$0.042 &    0.200 & $-$0.051 &
                                0.325 & $-$0.010 &    0.296 \\
 1.732 & \mbox{  }         &    0.233 & $-$0.028 &    0.178 & $-$0.049 &
                                0.316 & $-$0.001 &    0.269 \\
 2.000 & \mbox{  }         &    0.236 & $-$0.023 &    0.236 & $-$0.038 &
                                0.324 & $-$0.004 &    0.319 \\
\hline
 & &\multicolumn{2}{c}{Al-antisite}&\multicolumn{2}{c}{Fe-vacancy}&
    \multicolumn{2}{c}{Al-antisite}&Co-vacancy\\
 0.000 & \mbox{Al or Vac.} &    0.180 & $-$0.022 & $-$1.064 & $-$0.006 &
                                0.157 &    0     & $-$1.076 \\
 0.866 & \mbox{Al}         &    0.204 & $-$0.027 &    0.382 & $-$0.026 &
                                0.278 &    0     &    0.454 \\
 1.658 & \mbox{  }         &    0.250 & $-$0.034 &    0.215 & $-$0.046 &
                                0.343 &    0     &    0.313 \\
 1.000 & \mbox{TM}         & $-$0.282 &    0.392 & $-$0.204 &    0.838 &
                             $-$0.373 &    0     & $-$0.310 \\
 1.414 & \mbox{  }         & $-$0.242 &    0.643 & $-$0.282 &    0.635 &
                             $-$0.322 &    0     & $-$0.350 \\
 1.732 & \mbox{  }         & $-$0.234 &    0.758 & $-$0.299 &    0.603 &
                             $-$0.316 &    0     & $-$0.370 \\
 2.000 & \mbox{  }         & $-$0.202 &    1.199 & $-$0.206 &    1.454 &
                             $-$0.333 &    0     & $-$0.358 \\
\end{tabular}
\label{tab:Qm}
\end{table*}

It should be noted that in the present treatment we did not
account for a possible lattice relaxation of neighboring atoms
around a vacancy of any type, keeping in mind that such relaxation
cannot be reliably addressed in the ASA. To our knowledge, no such
relaxation studies (with the use of any full-potential scheme)
have yet been done for the systems in question. According to
Ref.~\onlinecite{Fu}, the relaxation around Pd vacancies in PdAl
is up to 2.5\%, and expectedly even less in FeAl, i.e. too small
to have a noticeable effect on magnetic moments.

As most interesting examples related to the development of
magnetic properties, the local DOS at different 
spheres around Fe and Co AS impurities in FeAl and CoAl are shown in
Fig.~\ref{Fe-AS_DOS} and \ref{Co-AS_DOS}, correspondingly.
The numerical data for charge transfer values and magnetic moments
inside atomic spheres of all defect systems studied are
presented in Table~\ref{tab:Qm}. These data correspond
to the lattice constant $a$=5.40 a.u. that is about the equilibrium
value for both ordered FeAl and CoAl. In the discussion that concerns
the bulk properties and defect formation energies (see below) we
refer to optimized equilibrium lattice spacings for defect systems, but
this is not so important for discussing partial charges.
The asymptotic value of the charge transfer within the perfect bulk
is roughly recovered in the crystal away from defects.
Some fluctuations in both charge transfer
and magnetic moments however prevail and are discussed below.

It is well known that the charge transfer is not uniquely determined,
since it depends on the choice of atomic spheres or cells
between which the charge flow occurs. Moreover, because of different
localization degree and spatial distribution of charge
in different solids it doesn't make sense to discuss the charge
transfer accompanying, e.g., the formation of a compound
from elemental constituents. This has been discussed at length
by Schulz and Davenport for $3d$ aluminides in Ref.~\onlinecite{SD93}.
With this in mind, we compare in Table~\ref{tab:Qm} the uncompensated
charges $Q$ in equal-sized atomic spheres at Al and TM sites
in perfectly ordered aluminides.
Charge-induced perturbations (Friedel oscillations) around the defects
also can be found in Table~\ref{tab:Qm}.
We emphasize that not the absolute values of $Q$, but their changes
over shells of neighbors to a defect are meaningful and of primary
interest here. The interesting feature seen in Table~\ref{tab:Qm}
is that both Fe-AS and Al-AS defects are less charged as compared
to respective bulk atoms of the ordered alloy. This is not surprising
because the electro-negativity of the nearest neighbors
is the same in both cases.

The general trends for the Fe--Al system could be already expected
based on the rigid-band model.
Since the electron transfer in the ordered alloy occurs from Al to TM,
the ways to lower the concentration of electrons
at the TM site are substitution of Al with a TM atom, or vacancies
of any kind. The Al-AS substitution, on the other hand,
is unfavorable for magnetism, as can be seen from Table~\ref{tab:Qm}.
The Fermi energy in ordered FeAl crosses the upper slope of a high
peak. According to the rigid band picture, the decrease
in the Al$\rightarrow$Fe charge transfer leads to an increase
of the local DOS value at the Fermi energy and thus favor the increase
in the magnetic moment in the framework of Stoner theory.

Consistently with this it is seen from Table~\ref{tab:Qm} that
the magnetic moment of Fe as an AS defect
(see also Fig.~\ref{Fe-AS_DOS}) or as a neighbor to the Al vacancy
is enhanced simultaneously with the decrease of the charge transfer
onto these sites. Because of such a locally changed charge transfer
the {\it local} DOS at the defect site is affected in a rather
complicated way, different from what one would expect
from rigid-band considerations.
The local DOS at the AS Fe, which is the central atom of the Fe$_9$
nearest-neighbor cluster, acquires certain similarity with the DOS of
pure bcc Fe with its half-filled minority-spin subband.
Furthermore the magnitude of the local magnetic
moment is quite close to that of bulk Fe. This is consistent with
the fact that practically no charge transfer occurs between
the central Fe atom and its nearest neighbors. From the experimental
studies\cite{DT} the local moment has been estimated to fall within the range
of 4 to 5.4 $\mu_{\mbox{\tiny B}}$, depending on the vacancy concentration.
According to our calculation, the cluster of nine Fe atoms
including the AS develops 8.3 $\mu_{\mbox{\tiny B}}$.
The local DOS of Fe atoms which are more distant
from the defect [Fe$(\frac{1}{2}\frac{1}{2}\frac{1}{2})$
and Fe$(\frac{3}{2}\frac{3}{2}\frac{3}{2})$ in Fig.~\ref{Fe-AS_DOS}]
become rapidly resembling that of Fe in ordered magnetic FeAl.
It is noteworthy that the spin-up subband is essentially filled,
so that any subsequent removal of electrons may take place primarily
from the minority-spin subband and thus increase the magnetic moment.
Therefore one should expect a clear correlation between
fluctuations of charge transfer and those of local magnetic moment over shells
of distant Fe neighbors. Apart from the Fe-AS atom, the pronounced
enhancement of the magnetic moment, compared to the average value
in the ordered alloy, occurs for Fe atoms which are first neighbors
to the vacancy at the Al site. 

{}From all four defects considered in the Co--Al system, only
the Co-AS develops a magnetic moment which is in agreement with previous
observations based on the KKR--Green's function
calculations of Stefanou {\it et al}.\cite{SZD} Quantitatively, our
results differ only slightly from those of Ref.~\onlinecite{SZD}:
our magnetic moment at the AS Co (1.53 $\mu_{\mbox{\tiny B}}$) is almost
the same as in pure elemental Co (1.72 $\mu_{\mbox{\tiny B}}$ experimentally;
1.55 $\mu_{\mbox{\tiny B}}$ as calculated, e.g., in Ref.~\onlinecite{SSKPD})
against 1.22 $\mu_{\mbox{\tiny B}}$ in Ref.~\onlinecite{SZD}.
The total moment at the cluster of nine Co atoms is 2.67 $\mu_{\mbox{\tiny B}}$
(2.06 $\mu_{\mbox{\tiny B}}$ in Ref.~\onlinecite{SZD}). This difference
is probably related to the fact that the magnetic moments are attributed
to space-filling atomic spheres, which we used in our case,
compared to non-overlapping muffin-tin spheres in the
KKR--Green's function approach.

An unusual feature in Table~\ref{tab:Qm} is that in two cases, 
namely for the Fe
vacancies and for the Al AS in FeAl, the Fe atoms which are
{\it most distant} from the defect exhibit quite large magnetic moments.
This is most probably an artefact
related to a yet too small supercell size. This artefact, which is
quite stable against the changes in the calculation setup
(more dense $k$-mesh in the BZ integration etc.), was not reported
in earlier LMTO supercell calculations by Gu and Fritsche.\cite{GF}
They used the same geometry of the 32-atoms supercell
(for AS defects) as we did and of course obtained
much similar results for the magnetic moments
in the vicinity of defects. Our explanation for the artefact is
the following: Magnetic moments within different shells of Fe
neighbors surrounding the defect exhibit long-range oscillations
around the bulk value of 0.76 $\mu_{\mbox{\tiny B}}$, as is seen
{}from Table~\ref{tab:Qm}. 
For the Fe atoms most distant from the vacancy at Fe and
from the Al AS, which are both situated at 2$a$ from the defect,
the oscillations just tend to be positive.
In our particular supercell geometry
there is only one of such an Fe atom per supercell, which at the same
time is the (fourth nearest) neighbor to six defect sites.
The constructive interference of these oscillations coming from
six defects is pinned at a single atom, resulting in an anomalously
high magnetic moment. In a real defect crystal with no ordering in
the mutual orientation of defects there is no physical reason
for such a high moment to be pinned at any site far from the defect.

\begin{figure}[t]
\epsfxsize=8.8cm
\centerline{\epsfbox{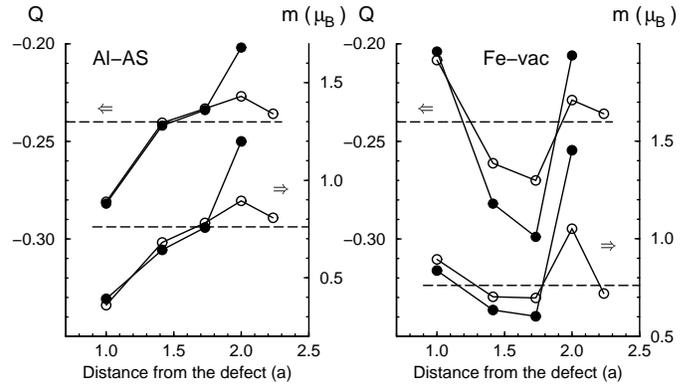}}
\vspace*{0.5cm}
\caption{Charge transfer (left scales)
and magnetic moments (right scales) over shells of Fe
neighbors to AS Al and Fe vacancy in FeAl. Dots -- calculation results
for 32-at. supercells; open circles -- for 64-at. supercells.
Dashed lines indicate the asymptotic values of charge transfer and
magnetic moment in ordered FeAl.}
\label{Oscil}
\end{figure}

In order to prove that the explanation given above is correct,
we have performed the calculations for supercells of increased
size (including 64 atoms) for two defect types in question.
The magnetic and charge distributions
over spheres are shown in Fig.~\ref{Oscil} in comparison with
the data for the 32-atom supercells, which are listed in Table~\ref{tab:Qm}.
One can see that,
while the anomalous magnetic moment at the fourth neighboring
Fe site is considerably reduced in a larger supercell,
the trends in the vicinity of defects are not as much affected.
Therefore the use of 32 atom-supercells for our subsequent analysis
is justified.  An additional observation seen from Fig.~\ref{Oscil}
is that charge fluctuations
and magnetic fluctuations over spheres behave almost identically.
This is due to the manifestation of the above mentioned feature that
local charge fluctuations affect essentially the minority-spin subband
and hence the net magnetic moment.

For CoAl, the Fermi energy is placed near the minimum
between two $d$-peaks and therefore leads to a loss of magnetic moment
in spite of decreasing of charge transfer due to the charge-induced
perturbations around the Co-vacancy,
Al-vacancy and Al-AS defects. Only the Co-AS defect provides a
sufficient decrease in the charge transfer in order to fulfill the
Stoner condition and to create the magnetic moment at this site.

\begin{table*}[hbt]
\caption{Cohesive properties (eV) at equilibrium lattice constants
$a$ (a.u.) of FeAl, CoAl and NiAl}
\begin{tabular}{ldddddd}
& \multicolumn{2}{c}{FeAl}
& \multicolumn{2}{c}{CoAl}
& \multicolumn{2}{c}{NiAl} \\
& Energy & $a$ & Energy & $a$ & Energy & $a$ \\
\hline
&\multicolumn{6}{c}{Formation energy} \\
\parbox{4.0cm}{
Experiment (Ref.~\protect\onlinecite{VC} \\*[-0.12cm]
and as cited in Ref.~\protect\onlinecite{SSKPD})}
& $-$0.26 & 5.409 & $-$0.56 & 5.408 & $-$0.64 & 5.456 \\
TB-LMTO (Ref.~\protect\onlinecite{SSKPD})
& $-$0.50 & 5.364 & $-$0.75 & 5.317 & $-$0.77 & 5.377 \\
Present work: Ordered B2
& $-$0.40 & 5.397 & $-$0.66 & 5.354 & $-$0.71 & 5.422 \\
Present work: Disordered
& $-$0.23 & 5.366 & $-$0.38 & 5.367 & $-$0.53 & 5.410 \\
&\multicolumn{6}{c}{Magnetic energy} \\
Ordered B2 & $-$0.02 & & ---  & & --- \\
Disordered & $-$0.05 & & 0.00 & & --- \\
\end{tabular}
\label{tab:cohesive}
\end{table*}

It is necessary to point out that planar defects, like the surface,
may be expected to affect the magnetism in a similar way as
point defects do. In particular, FLAPW 7-layer slab calculations,
performed in order to simulate the surfaces
of FeAl and CoAl compounds,\cite{OFC} resulted in a magnetic ordering
for FeAl(100) and CoAl(100) surfaces, while NiAl(100) was magnetically dead.
The Fe surface atom with a magnetic moment of 2.57 $\mu_{\mbox{\tiny B}}$
has up to 0.25 
electrons less than the next iron layer atom with a moment
of 0.67 $\mu_{\mbox{\tiny B}}$ -- this is the same trend, but more
pronounced, as in our calculation for a Fe atom neighboring the ``cavity''
of the Al vacancy. In the case of CoAl, the surface atom loses
about 0.3 electrons as compared to the atom in the next Co layer,
which results in the enhancement of the surface Co moment
to 1.12 $\mu_{\mbox{\tiny B}}$.

\begin{figure}[bh]
\epsfxsize=8.2cm
\centerline{\epsfbox{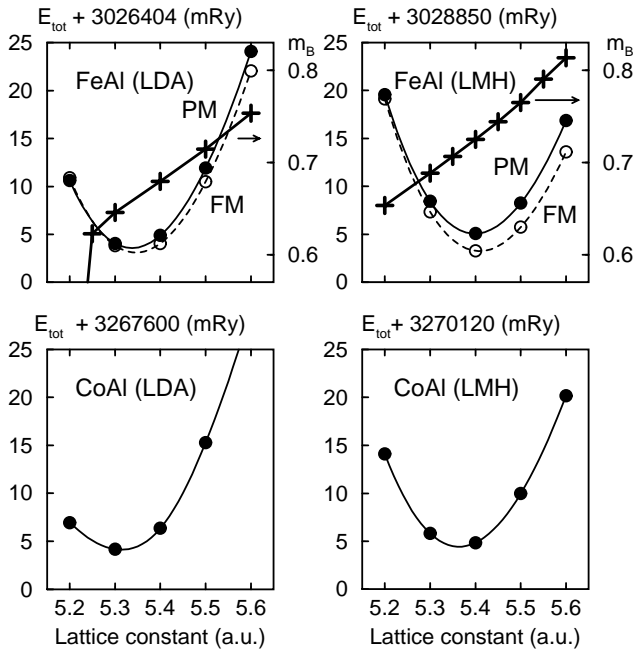}}
\vspace*{0.5cm}
\caption{Total energy (dots, left scale) per formula unit
vs. lattice constant in ordered FeAl and
CoAl as calculated by TB-LMTO in the LDA and with non-local
corrections (LMH).}
\label{LMTO_E_of_a}
\end{figure}

The changes in the total energies due to the variations
of the lattice constants exhibit the expected parabolic behavior.
This is shown in Fig.~\ref{LMTO_E_of_a},
where the results for two different exchange-correlation potentials
are presented. Table~\ref{tab:cohesive} includes the results 
obtained by a parabolic fitting
of the total energy data. As pointed out in
Sec.~\ref{methods}, the data related to ordered compounds, as well as
the 

\noindent
defect formation energies, have been obtained with
non-local corrections to the exchange-correlation, whereas the data
for disordered alloys (discussed in more detail below)
result from the calculations within the LDA.
The calculated equilibrium lattice constants are consistently
underestimated by about 1\%, with somehow better agreement
to the experimental data in the case where the non-local corrections
have been included.
The comparison with the results of earlier TB-LMTO calculations\cite{SSKPD}
for ordered TM aluminides, obtained within the same formalism but in the LDA,
reveals that the non-local corrections to the LDA produce indeed
a systematic improvement of all cohesive properties for the
alloys under consideration.

The cohesive properties have been calculated at equilibrium volumes
and are compared with the experimental data available.
The experimental data in Table~\ref{tab:cohesive} are from thermochemical
formation energy measurements; the total energy of the constituent
solids are taken at the equilibrium volume in their respective
ground-state structures, i.e. fcc Al and Ni, hcp Co and bcc Fe
(ferromagnetic Fe, Co and Ni), calculated with the same setup
as used for alloys. The total energies in the TB-LMTO calculation
have been also minimized for the defect supercell systems,
thus accounting for a uniform lattice dilatation around defects.
The corresponding lattice constants are shown in Table~\ref{tab:a_scell}.

\begin{table}[b]
\caption{Equilibrium lattice constants $a$ (a.u.)
from TB-LMTO supercell calculations}
\begin{tabular}{cdddd}
& TM vacancy & Al vacancy & TM antisite & Al antisite \\
\hline
FeAl & 5.404 & 5.401 & 5.387 & 5.424 \\
CoAl & 5.355 & 5.337 & 5.341 & 5.381 \\
\end{tabular}
\label{tab:a_scell}
\end{table}

It is noteworthy to mention that the calculated heats of formation
come out systematically overestimated for all ordered compounds,
but to a less extent in the present work than in earlier
TB-LMTO calculations\cite{SSKPD} with local exchange-correlation.
On the other side, they are systematically underestimated
by about the same margin in the KKR-CPA calculations (see next Section).
Apparently, the properties of real systems fall between the two
extremities of complete substitutional disorder and complete ordering
with point defects, which are addressed in the present study.

In our calculations ordered CoAl and NiAl fail to develop
any magnetic ordered state in agreement with all previous calculations,
while FeAl has both ferromagnetic (FM) and nonmagnetic solutions.
The earlier LMTO\cite{MOJF} and TB-LMTO\cite{SSKPD} studies
found the nonmagnetic solution as the ground state within the LDA;
our calculation with the LDA puts the magnetic solution minutely
lower in energy, actually one cannot reliably claim this
(see Fig.~\ref{LMTO_E_of_a}). With
non-local corrections included to the exchange-correlation,
the ferromagnetic ground state clearly gains in energy
with respect to the nonmagnetic state.
The values of the magnetic moment (per cell) for different values
of lattice constant are also shown in Fig.~\ref{LMTO_E_of_a}. As is consistent
with other studies on magnetic systems, the non-local exchange-correlation
results in somehow larger values of magnetic moments.
It should be noted also that ASW fixed-spin-moment calculations\cite{MM}
provided, in addition, the antiferromagnetic (AFM) state with an energy
difference between FM and AFM states of only a tenths of a mRy.
Since three possible ground states have very small energy differences,
one can conclude that the formation of a spin-glass state seems plausible
in the Fe--Al system. The existence of the spin-glass state was indeed
detected in Ref.~\onlinecite{SW}.

\section{Electronic structure of disordered alloys}
\begin{figure}[ht]
\epsfxsize=9.0cm
\centerline{\epsfbox{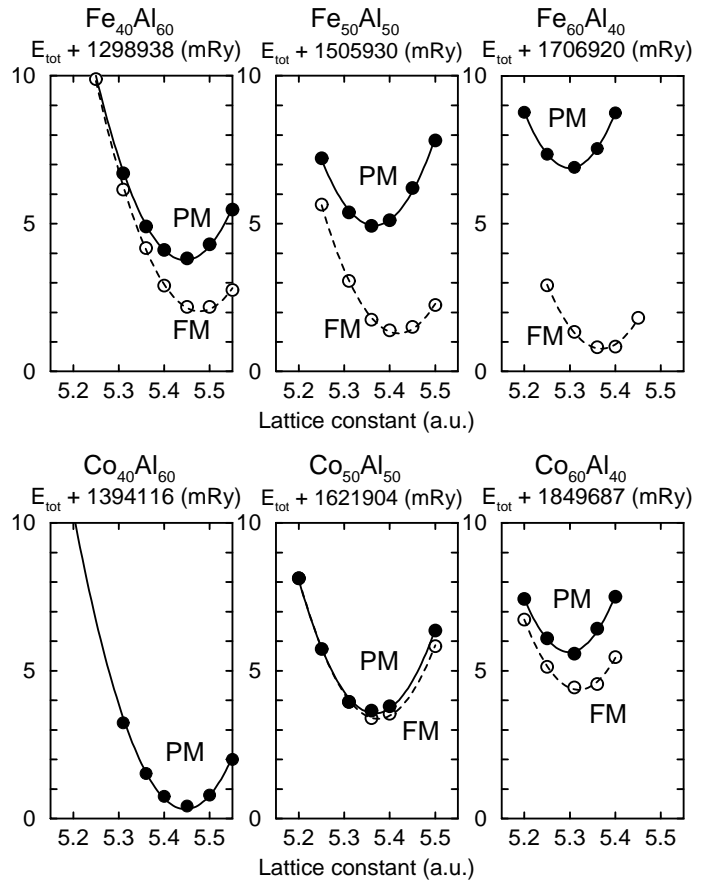}}
\vspace*{0.5cm}
\caption{Total energy (per atom) vs. lattice constant
in disordered Fe--Al and
Co--Al for several concentrations as calculated by KKR-CPA,
for paramagnetic (PM) and ferromagnetic (FM) solutions.}
\label{CPA_E_of_a}
\end{figure}

While the supercell approach aims at studying point defects
in otherwise perfect alloys, the limiting case of the ultimate
substitutional disorder can be treated by the KKR-CPA approach.
Here an atom of each species in the lattice with the probability
equal to its concentration can be considered as an impurity
in the effective medium, and with complementary probability --
as the AS defect. We have performed such calculations for Fe$_x$Al$_{1-x}$,
Co$_x$Al$_{1-x}$ and Ni$_x$Al$_{1-x}$ alloys in the range of compositions
$x=0.4$ to $0.6$. From the series of KKR-CPA calculations
for different alloy compositions and different volumes per atom
we have determined the equilibrium lattice parameters. Their values
for $x=0.5$ are also listed in Table~\ref{tab:cohesive}.
In Fig.~\ref{CPA_E_of_a} we show the results of these calculations
for Co--Al and Fe--Al systems for both ferromagnetic and
nonmagnetic solutions. The magnetic solution is missing
in Co$_x$Al$_{1-x}$ only for $x=0.4$.
For Ni$_x$Al$_{1-x}$ no magnetic solution has been found
in the whole range of concentrations, which have been studied

\noindent
in this contribution.

As it is shown in Fig.~\ref{CPA_E_of_a}
the Co$_{0.5}$Al$_{0.5}$ alloy exhibits
the ferromagnetic solution, in contrast to the ordered CoAl compound
(see Fig.~\ref{LMTO_E_of_a}).
The total energy difference between ferromagnetic
and nonmagnetic ground states is however extremely small,
as is listed in Table~\ref{tab:cohesive}. The magnetic moment at the Co site
is 0.47 $\mu_{\mbox{\tiny B}}$ near the transition point to
the nonmagnetic state. The ferromagnetic state disappears
under pressure but becomes clearly the favorable one for
increased volumes in this system.

\begin{figure}[hbt]
\epsfxsize=8.0cm
\centerline{\epsfbox{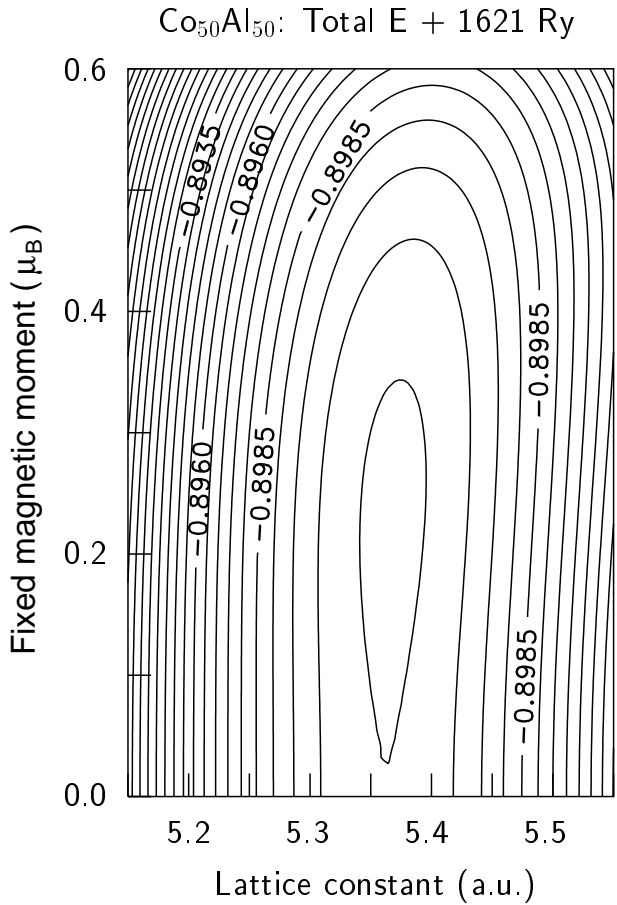}}
\vspace*{0.5cm}
\caption{Total energy vs. lattice constant and magnetic moment
calculated by FSM KKR-CPA for disordered Co$_{0.5}$Al$_{0.5}$.}
\label{Fixmom}
\end{figure}

In order to investigate the type of magnetic instability,
we have performed the FSM analysis for the
Co$_{0.5}$Al$_{0.5}$ alloy, with the total energy calculated
as a function of fixed magnetic moment and lattice constant.
The results of these calculations shown in Fig.~\ref{Fixmom}
indicate the presence of a saddle point between magnetic and
nonmagnetic solutions. It means that the phase transition between
ferromagnetic and nonmagnetic states occurs as a first-order transition.
At the moment, this is only a theoretical prediction, since
the fully disordered CoAl alloy was not obtained
experimentally and, moreover,
the magnetic transition is expected to occur for a somehow
expanded lattice. However, this transition can be seen 

\noindent
as a kind of pressure-induced transition in the case of
Co-rich disordered alloys,
as follows from Fig.~\ref{CPA_E_of_a}, and it has been really found
at low temperatures in the Co$_{0.6}$Al$_{0.4}$ alloy.\cite{SCF}

For Fe--Al disordered alloys, the ferromagnetic solutions
exist over the whole range of composition $0.4<x<0.75$, which has been covered
in the present work. The results are shown in Fig.~\ref{CPA_E_of_a}.
The energy difference between ferromagnetic and nonmagnetic solutions
increases for Fe-rich disordered alloys, and for the
Fe$_{0.5}$Al$_{0.5}$ alloy we obtain the ferromagnetic ground state
already within the LDA approach (see, also, Table~\ref{tab:CPA_ene}).
Only for the case $x<0.4$ the nonmagnetic solution
can be more stable. It is interesting to point out
that for both Co--Al and Fe--Al disordered alloys the minimum
of the formation energy coincides with the stoichiometric composition.
On the contrary, for the Ni--Al alloy the minimum is shifted
to a higher concentration of the transition metal.

\begin{table}
\narrowtext
\caption{Disordered alloys formation energies (eV)}
\begin{tabular}{ldddd}
Alloy & $x$ & electron/atom & ferro & para\\
\hline
Fe$_x$Al$_{1-x}$ & 0.4  & 5.00 & $-$0.25 & $-$0.23 \\
                 & 0.5  & 5.50 & $-$0.28 & $-$0.23 \\
                 & 0.55 & 5.75 & $-$0.28 &         \\
                 & 0.6  & 6.00 & $-$0.28 & $-$0.20 \\
                 & 0.65 & 6.25 & $-$0.26 &         \\
                 & 0.7  & 6.50 & $-$0.25 &         \\
                 & 0.75 & 6.75 & $-$0.22 &         \\
Co$_x$Al$_{1-x}$ & 0.4  & 5.40 &         & $-$0.36 \\
                 & 0.5  & 6.00 & $-$0.38 & $-$0.38 \\
                 & 0.6  & 6.60 & $-$0.36 & $-$0.34 \\
Ni$_x$Al$_{1-x}$ & 0.4  & 5.80 &         & $-$0.45 \\
                 & 0.5  & 6.50 &         & $-$0.53 \\
                 & 0.6  & 7.20 &         & $-$0.55 \\
\end{tabular}
\label{tab:CPA_ene}
\end{table}

\begin{figure}[th]
\epsfxsize=8.2cm
\centerline{\epsfbox{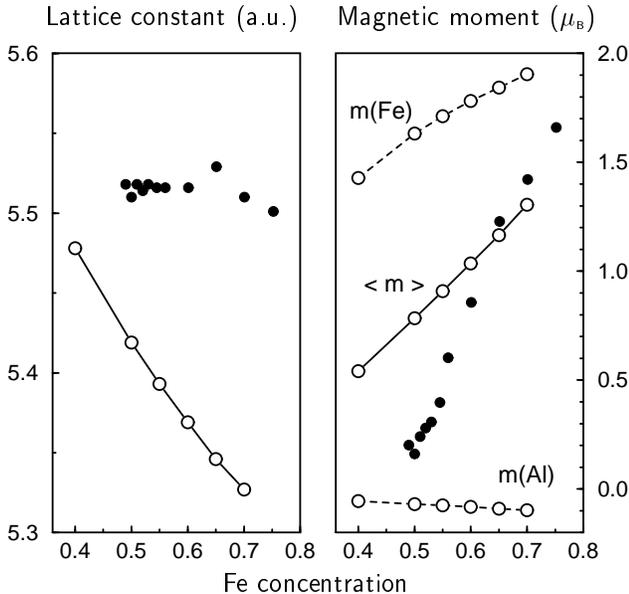}}
\vspace*{0.5cm}
\caption{Theoretical KKR-CPA (open circles) and experimental (dots)
concentration dependence of the lattice constant and the magnetic moment
in the Fe--Al alloys.}
\label{CPA_M_of_x}
\end{figure}

The Fe--Al alloys, annealed from high temperatures in order to obtain
the perfect disordered state, show the magnetization in a broad range
of compositions.\cite{LB} We compare these experimental values with
the calculated KKR-CPA magnetization per unit cell in
Fig.~\ref{CPA_M_of_x},
where also the local magnetic moments for Fe and Al atoms are given.
Actually, the calculated magnetization decreases slower with
the Al concentration than it does in the experimental findings.
This can be explained by the change of disordering degree
with concentration in the actual samples.
The comparison between the experimental and the calculated dependencies
of the lattice constant supports this possibility (see Fig.~\ref{CPA_M_of_x}).
The experimental dependence of the lattice constant on composition
can be considered as an averaging over local variations of spacing
around randomly distributed defects, as affected, e.g., by
the heat treatment. The KKR-CPA calculation, on the contrary,
gives the equilibrium volume as a smooth monotonic function
of the iron concentration. As usual, the magnetization is a function
of the equilibrium volume, and an extended lattice
provides a larger value of the magnetization.
In order to get more insight into the volume trends on disordering,
the data on equilibrium lattice constants for defect systems,
as estimated from our supercell calculations (see Table~\ref{tab:a_scell}),
may be of some use. We did not allow for local (breathing)
relaxations of neighbors around the defect because
such energy evaluations might be not sufficiently reliable
in a calculation done with the ASA. The trends in the uniform lattice
expansion, simulating in this case the effect of a lattice dilatation
around defects, are usually quite robust and reasonably obtainable
in the ASA. The equilibrium lattice constant
for a system with an Al-AS defect is 5.424 a.u., whereas that with an Fe-AS
defect is 5.387 a.u.
The only possibility to understand the experimental trend seen
in Fig.~\ref{CPA_M_of_x}, i.e. that the lattice constant is
relatively independent on the Fe concentration, -- is to assume
the existence of Al antisite defects even at elevated Fe concentration.

\begin{figure}[ht]
\epsfxsize=9.2cm
\hspace*{-0.7cm}\epsfbox{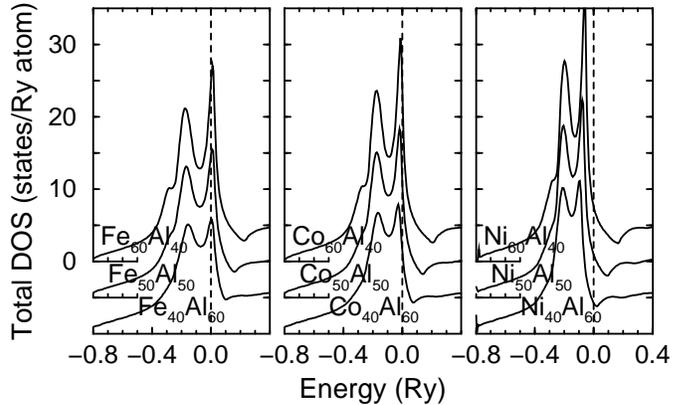}
\vspace*{-0.7cm}
\caption{Total DOS for several concentrations of disordered Fe--Al, Co--Al
and Ni--Al alloys as calculated by KKR-CPA.}
\label{All_CPA_DOS}
\end{figure}

It is very useful to compare the DOS for the disordered alloys series
Ni--Al, Co--Al and Fe--Al as we do in Fig.~\ref{Single_DOS}
for ordered compounds. This comparison is shown in Fig.~\ref{All_CPA_DOS}.
As becomes evident from our results,
the rigid-band model applies better for any fixed TM-Al composition
with different TM components than for the same alloy with
different concentrations. In particular, the low-energy structure at
about $-$0.3 eV is present for Ni--Al, Co--Al and Fe--Al alloys with 60\%
of TM but it disappears for the alloys with smaller TM concentrations.
Also, the sharpness of the peak at the Fermi energy
remarkably changes with the concentration.

The main features of the DOS profiles for ordered compounds are assigned
by Zou and Fu,\cite{ZF} in the order of increasing of energy,
to the bonding, nonbonding (the peak just below $E_F$ in NiAl)
and antibonding states (Fig.~\ref{Single_DOS}).
A 'pseudogap' separating the bonding and nonbonding states remains
well defined in all DOS calculations for ordered compounds.
As is shown in Fig.~\ref{All_CPA_DOS}, the nonbonding structure
completely disappears for disordered alloys.
The tendency for the disappearance of the
nonbonding peak can be already traced in Figs. \ref{Fe-AS_DOS} and
\ref{Co-AS_DOS} for the AS local DOS,
but only the complete disorder leads to the total smearing out
of this DOS feature. This effect is favorable for the magnetism,
which develops in Co--Al disordered alloys. On the other hand,
the disorder reduces the formation energy (see Table~\ref{tab:cohesive})
because the Al to TM charge transfer requires that a late TM atom
has as many Al atoms as its nearest neighbors as possible,
in order to facilitate the charge transfer and the bonding hybridization.

\section{Conclusion}
We have shown in this paper that the onset of magnetism in the Ni, Co and Fe
aluminides is closely related to the defect structure of these compounds.
Among the perfectly ordered 1:1 compounds, only FeAl retains
a magnetic moment of 0.76 $\mu_{\mbox{\tiny B}}$ per atom.
In CoAl and NiAl the magnetic moments are totally quenched
due to the $d$-band population.
However, in the case of TM-AS defects the value of the magnetic moment
for a Fe defect is the same as for a pure bcc iron atom.
A large magnetic moment also appears for the Co-AS defect.
All nearly disordered Fe--Al alloys exhibit a ferromagnetic ordering
over the range of compositions 75 at.\% to 40 at.\% of Fe,
i.e. the existence range of the B2 phase.
In Co--Al disordered alloys, the onset of ferromagnetism
occurs only for a Co concentration larger than 50 at.\%.
The Ni--Al alloys do not show any magnetic ordering.

The Friedel oscillations around the defects are favorable
for the enhancement of magnetic moments in the ferromagnetic FeAl matrix.
It means that any defect creates an enhancement of the magnetic moment
on neighboring Fe atoms, as charge transfer to these atoms decreases.
In the paramagnetic Co--Al matrix, however, these oscillations are only
sufficient for creating magnetic moments on Co sites,
which are directly neighbors to the Co AS impurities.

\section*{Acknowledgments}
This work was supported by the Deutsche Forschungsgemeinschaft.
N.I.K. thanks the University of Osnabr\"uck
for the kind hospitality during his stay there.
Useful comments by J.~Kudrnovsk\'y are appreciated.

\end{document}